# Title: Solving the puzzle of tropical cyclone track bifurcation

**Authors:** Wei Zou[1], Jizhou Li[2], Xurui He[1], Lei Liu[1], Pinaki Chakraborty[3], Lin Li[1]*

**Affiliations:**

[1]Department of Mechanics, Sichuan University; Chengdu, China.
[2]Research Institute for Electronic Science, Hokkaido University; Sapporo, Japan.
[3]Fluid Mechanics Unit, Okinawa Institute of Science and Technology Graduate University; Onna, Japan.
[*]Corresponding author. Email: lin.li@scu.edu.cn

**Abstract:** Tropical cyclones whose predicted tracks bifurcate are among the most dreaded storms in disaster planning. Understanding when and why these bifurcations occur has been a long-standing puzzle in track forecasting. We address this issue by introducing the concept of "fate boundaries" hidden in the steering flow. Storms located on opposite sides of these boundaries follow dramatically different tracks. By analyzing 1,179 Northern Hemisphere storms over 2011–2025, we identified 61 with a strong signature of bifurcation. In every instance, fate boundaries explained why the ensemble tracks split. Additionally, they can reveal bifurcations days before they appear in any operational ensemble. Because fate boundaries often separate devastating from nearby benign tracks, they mark when and where minimal interventions could, in principle, redirect a storm.

## Main Text:

Tropical cyclone track forecasting relies on ensembles (*1*, *2*). In an ensemble, tropical cyclones are initialized at slightly different positions (representing positional uncertainty at that instant) (*1*, *3*), and these storms are tracked forward in time, steered by the large-scale flow (*4*). Typically, the resulting tracks fan out smoothly (Fig. 1A). In some cases, however, the tracks bifurcate into distinct branches (*5*) (Fig. 1B). Some of the branches can split further, creating a tree-like structure (Fig. 1C). Such cases pose an acute dilemma over which regions to warn and evacuate (*6*). Yet, why and when the bifurcation occurs has remained a longstanding puzzle (*7*–*9*).

This gap in understanding has a serious consequence for disaster planning. Ensembles make predictions based on sampling (*1*, *2*, *10*). Without *a priori* guidance on when a bifurcation could occur, ensembles can completely fail to capture it. A stark example is Khanun (2023) (*11*). All members of the early ensemble predicted it would strike China; instead, it stalled offshore and veered sharply toward Japan (Fig. 1D). The Japan-bound branch did not emerge until shortly before the turn. The late emergence of this branch forced preparedness efforts to be redirected at short notice. The problem is common to both traditional numerical models (Fig. 1D) and AI models, such as Google DeepMind's WeatherNext 3 (*12*, *13*) (fig. S1). There are two possible approaches to address this problem. One is to substantially increase the ensemble size to better sample possible tracks (*14*, *15*). However, this can be computationally expensive, and there is no guarantee that all branches will be sampled (*14*–*17*), even for very large ensembles. The other is to develop a physics-based method that can predict *a priori* when bifurcation would occur.

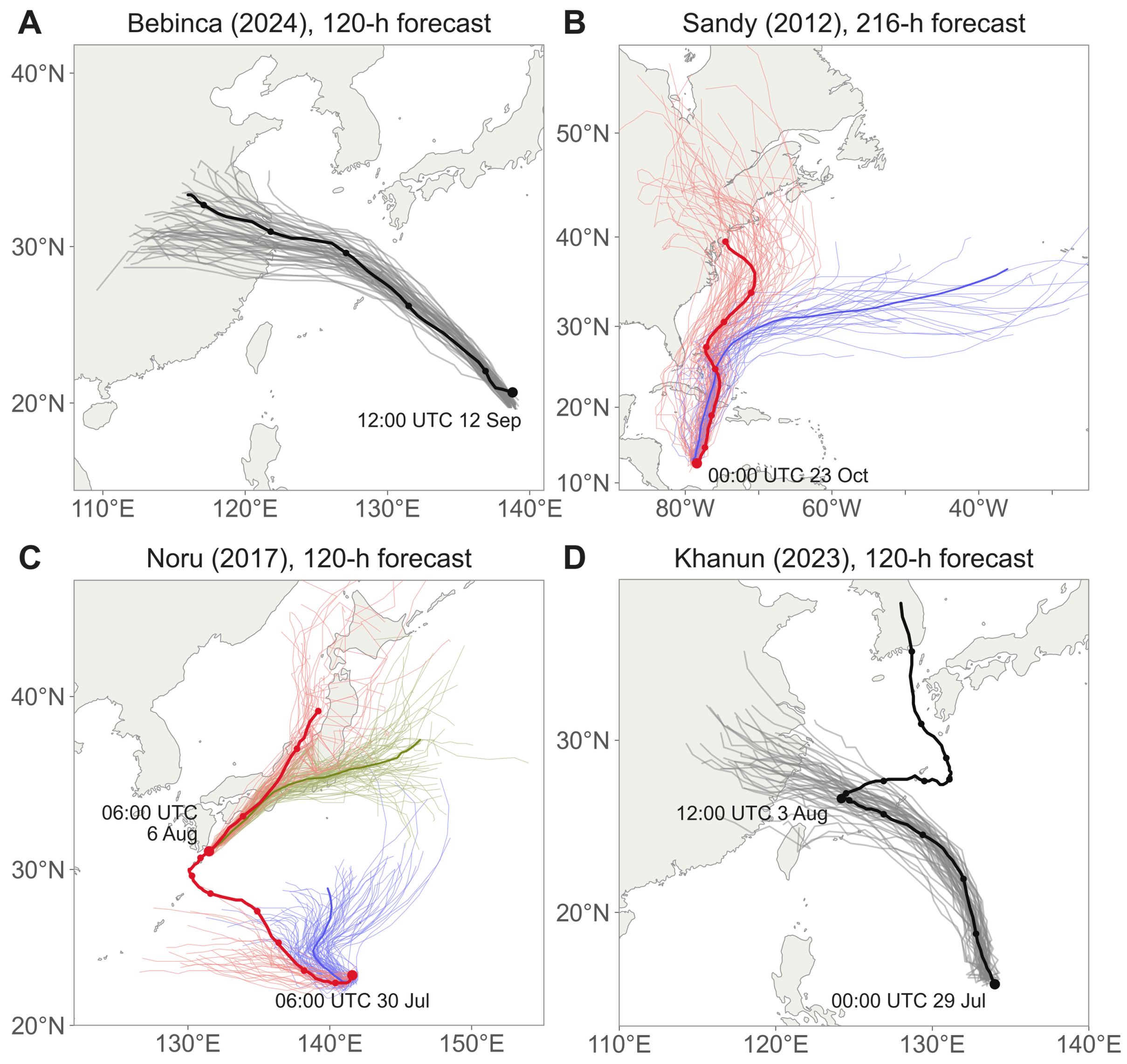

**Fig. 1. Non-bifurcating and bifurcating ensemble tracks.** Thin lines are ensemble members; thick lines with dots are the observed track, and thick lines without dots are the mean track of each alternative branch. Where the ensemble bifurcates, members are colored by branch. (**A**) Bebinca (2024). (**B**) Sandy (2012). (**C**) Noru (2017). (**D**) Khanun (2023).

The prevailing physics-based approach to predict bifurcation is based on the "col" framework (*7*, *8*, *18*), which is rooted in how a tropical cyclone moves. Because a tropical cyclone is embedded in, and steered by, the large-scale flow, it is usually straightforward to infer its direction of motion from that flow (*4*). However, when two competing steering circulations are present, their steering effects can nearly cancel in the region between them, creating a weak-steering region known as a "col" (*7*, *8*, *18*). There, small differences or perturbations can send the storm in very different directions, potentially giving rise to distinct branches. [Another physics-based method, which is closely related to the col, is the axis of contraction (*19*); we discuss that in fig. S2.]

**Fate boundaries in a minimal model**

Can the col framework accurately predict bifurcation? We interrogate this issue in an idealized setting in which the initial positions of storms can be precisely controlled. Inspired by Khanun, we construct a minimal model in which we represent two anticyclones that exerted competing steering on Khanun (Fig. 2A) (*11*) as two point vortices. These vortices are fixed in position but slowly vary in strength with time, mirroring their actual counterparts [supplementary materials (SM), materials and methods (*20*)]. We release three pairs of initially adjacent tracers in the resulting flow and track their time evolution (Fig. 2B). Seeded within the col, tracers in pair 1 split abruptly. One drifts west, while the other hooks sharply east, in a turn reminiscent of Khanun's. This appears to support the col framework. However, tracers in pair 2, also seeded within the col, remain close throughout, whereas those in pair 3, seeded outside the col, split. Starting within the col is therefore neither sufficient (pair 2) nor necessary (pair 3) for bifurcation.

The above results show that the col framework is unable to predict bifurcation. Here we take a different approach. We draw inspiration from dynamical systems studies of atmospheric pollutant transport and oceanic oil spills (*21*, *22*), where the flow carrying the material has a skeleton known as repelling Lagrangian coherent structures (LCS) (*23*, *24*). These structures act as hidden boundaries between trajectories heading in markedly different directions. Importantly, these structures reflect the evolution of the flow rather than a snapshot, thus offering a Lagrangian perspective (in contrast to the instantaneous picture underlying the col framework). In this work, we test whether such structures can predict tropical cyclone track bifurcation. To our knowledge, this is the first application of LCS to track forecasting.

To locate LCS, we compute the finite-time Lyapunov exponent (FTLE) (*23*, *25*), which measures how fast initially adjacent tracers separate over a finite time window (Fig. 2C). Notably, the region with high values of FTLE is confined to a narrow, almost one-dimensional band. This band threads precisely between the two tracers of each pair that splits (pairs 1 and 3) and bypasses the pair that does not (pair 2). In dynamical-systems terms, the ridge (the locus of cross-band maxima) of such bands marks a repelling LCS, across which nearby trajectories separate rapidly. In track forecasting, this structure is in effect a "fate boundary": tracers straddling it, however close initially, are fated to move along trajectories toward distinctly different destinations. Beyond the three illustrative pairs, we seed tracers throughout the domain. Every pair of tracers straddling the fate boundary splits, and no other pair does (Fig. 2D). Straddling the fate boundary is thus both necessary and sufficient for bifurcation in this model.

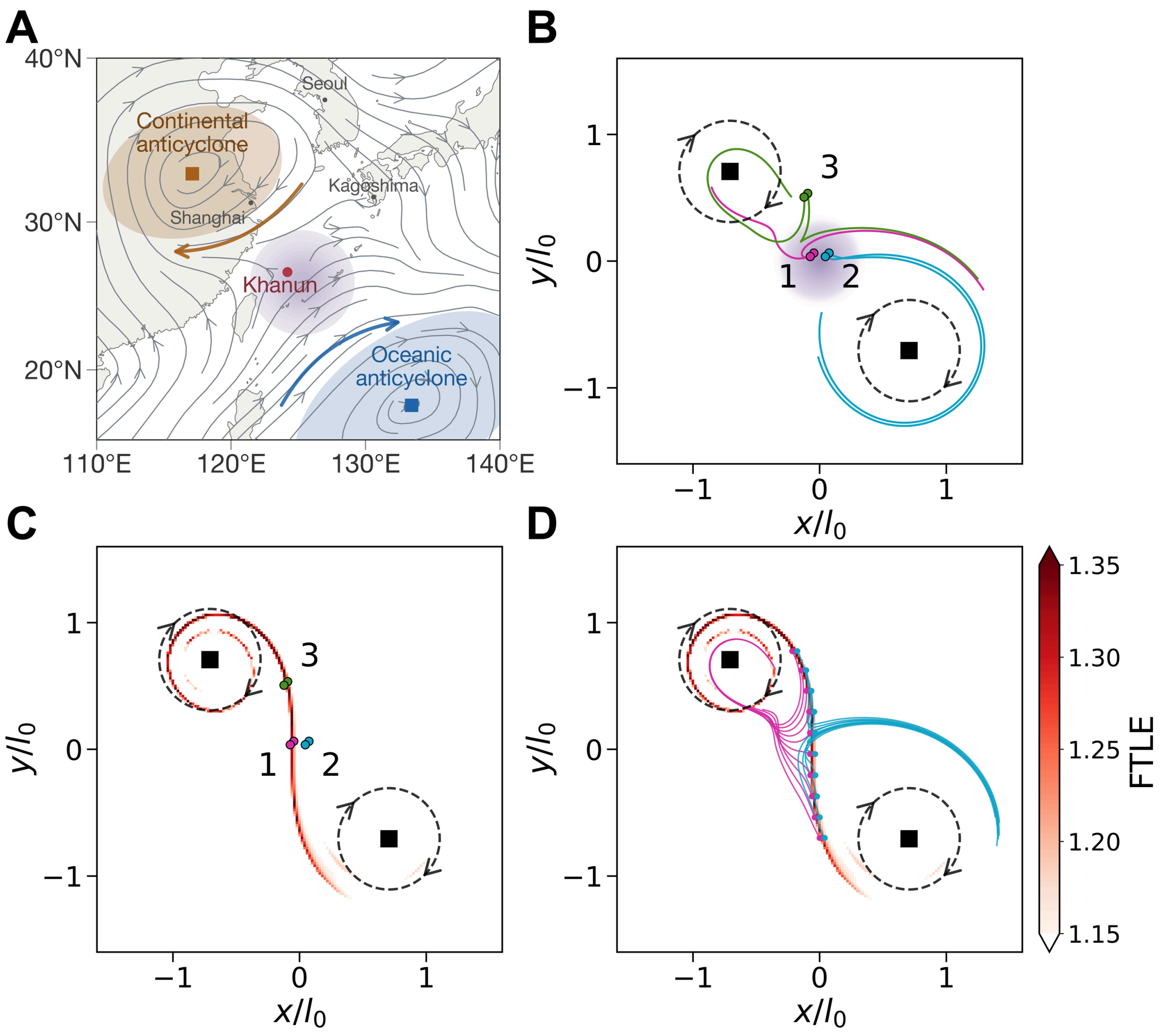

**Fig. 2. Fate boundaries in the minimal model.** (**A**) Steering flow at Khanun's turn, 12:00 UTC 3 August 2023. Squares and shading mark the centers and extents of the continental and oceanic anticyclones; purple shading marks the col region. (**B** to **D**) The minimal model: two point vortices with clockwise (anticyclonic) circulation, fixed in position but with time-varying strengths; lengths and time are dimensionless (SM, materials and methods). (**B**) Three tracer pairs and their trajectories; purple shading marks the col region around the stagnation point at t = 0. (**C**) FTLE field, along with the starting positions of the three tracer pairs from (B); the ridge of the FTLE field marks the fate boundary. (**D**) Ten tracer pairs seeded along the fate boundary.

**Fate boundaries in the real atmosphere**

In the minimal model, the steering flow is highly idealized, and the FTLE field cleanly captures the fate boundary. Real atmospheric flows, however, are far more complex. Can the FTLE field still be used to identify fate boundaries and predict bifurcation? To test this, we turn to the actual steering flow that carried Khanun.

We use the ERA5 reanalysis (*26*), which is widely regarded as the best available estimate of the true atmosphere. In previous LCS studies of atmospheric material transport (*27*), tracers represent air parcels carried by the total wind, including the storm's wind. For tracers to represent storm motion, we must remove the storm's own wind. Otherwise, tracers initialized around the storm center would be driven by the storm's circulation rather than follow its track. We therefore remove the storm's circulation and then average the remaining wind over the storm's depth to obtain a two-dimensional steering flow. Because storms also undergo "beta drift" relative to the steering flow, we advect the tracers with the beta-advection model (*28*), which incorporates this additional motion (SM, materials and methods). Each tracer thus represents a possible storm trajectory. We release tracers at 00:00 UTC on 1 August 2023, two and a half days before Khanun's turn, and compute FTLE from their five-day trajectories.

The atmosphere yields a field unmistakably busier than the minimal model (Fig. 3A). Filaments of high FTLE crowd the plane, as if fate boundaries were everywhere. Much of the crowding, however, is irrelevant for bifurcation. The standard FTLE registers tracer separation, including separation by shear (*29*, *30*), which pulls tracers apart along similar paths toward the same geographical region. Such separation does not form a bifurcation. To focus on track bifurcation, we introduce a direction-sensitive counterpart of the FTLE (hereafter "directional FTLE"; fig. S3; SM, materials and methods). It retains divergence toward different directions and discards separation that preserves a common direction (Fig. 3, A and B).

The resulting field is markedly cleaner (Fig. 3B). Two ridges stand out in Khanun's neighborhood: $\gamma_1$ to the storm's west and $\gamma_2$ to its east. We first examine $\gamma_1$. We release pairs of tracers straddling $\gamma_1$ (Fig. 3C). Notably, tracers released west of $\gamma_1$ strike China, while tracers released east of $\gamma_1$ recurve toward Japan. Pairs that begin 50 km apart end, five days later, in different countries, parted by the full width of the East China Sea (~1,000 km). That Khanun lies on the eastern side of $\gamma_1$ means the storm is fated to recurve toward Japan. Yet, Khanun's proximity to $\gamma_1$ also means that even small member-to-member differences in the ensemble can shift the storm across $\gamma_1$, headed toward China.

We next examine $\gamma_2$. Seeding tracer pairs across $\gamma_2$ (Fig. 3D), we find that the side nearer Khanun's position routes tracers to mainland Japan, the recurving path already revealed by $\gamma_1$. The far side, however, routes tracers neither to China nor to mainland Japan, but out over the open Pacific. Curiously, this fate appeared in none of the ensemble members (fig. S4).

To examine the robustness of the concept of fate boundaries to explain bifurcation, we now consider a much broader set of bifurcating tracks. We apply an objective clustering algorithm to all 1,179 Northern Hemisphere storms from 2011 to 2025 (SM, materials and methods). This yields 61 cases with a strong signature of bifurcation in their predicted tracks. In every case, a fate boundary lies near the storm, typically within 200 km (fig. S5), and seeding across it reproduces the bifurcating structure of the ensemble (Fig. 4 and figs. S6 to S8). The fate boundary thus explains why the ensemble tracks split. Moreover, we identify four storms for which the forecast ensemble misses an entire branch (fig. S9).

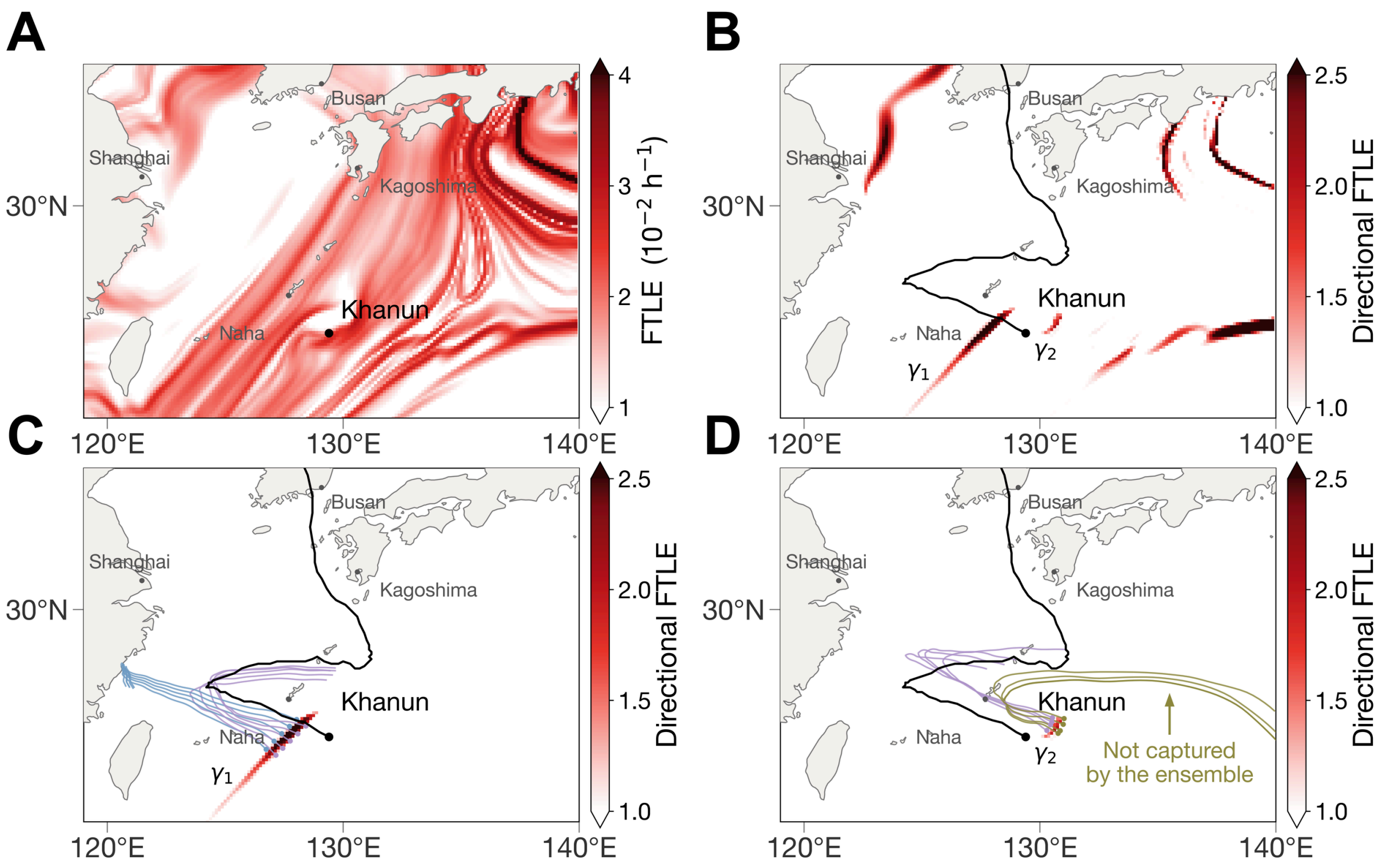

**Fig. 3. Fate boundaries for Khanun.** All panels are computed from tracers advected forward for 120 h from 00:00 UTC 1 August 2023 under the beta-advection model (SM, materials and methods). The black dot marks Khanun's position at 00:00 UTC 1 August 2023; in (B) to (D), the black line marks its observed track. (**A**) Standard FTLE. (**B** to **D**) Directional FTLE; the two ridges near Khanun are labeled $\gamma_1$ and $\gamma_2$. (**C** and **D**) Tracer pairs seeded 25 km to either side of $\gamma_1$ (C) and $\gamma_2$ (D).

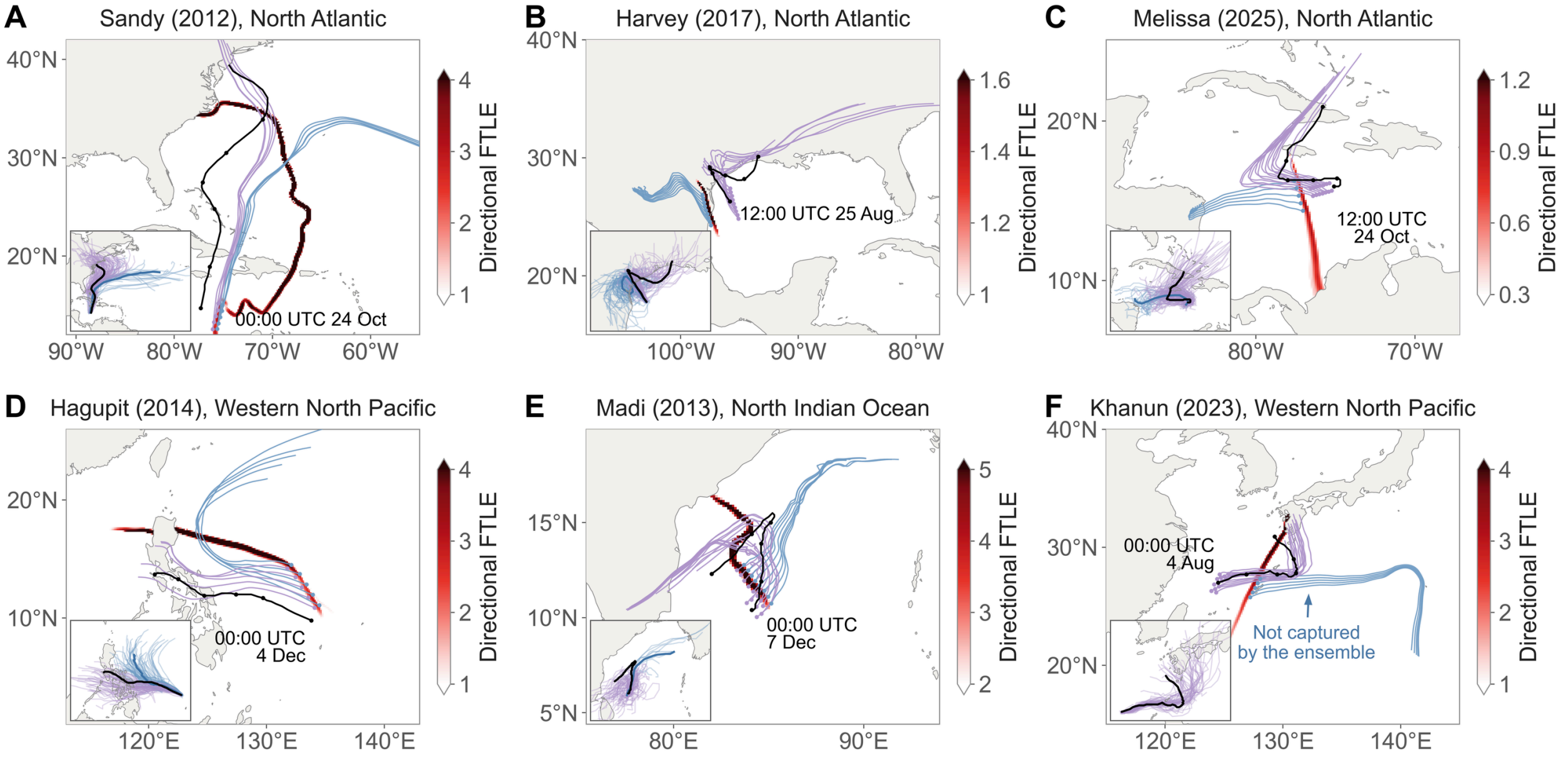

**Fig. 4. Fate boundaries in six storms.** Directional FTLE (red shading) and tracer trajectories for (**A**) Sandy (2012), (**B**) Harvey (2017), (**C**) Melissa (2025), (**D**) Hagupit (2014), (**E**) Madi (2013), and (**F**) Khanun (2023), computed as in Fig. 3. Only the fate boundary relevant to the storm's bifurcation is shown; blue and purple tracers are seeded on either side of it. Insets show the operational ensemble forecasts; we mark the two branches by different colors. Black lines show observed tracks. Note that in (F), Khanun's turn toward the Korean Peninsula at 00:00 UTC 7 August 2023 is also associated with a fate boundary; the alternative fate, toward the open Pacific, is absent from operational ensemble forecasts (inset).

**Fate boundaries in the forecast atmosphere**

We next turn from reanalysis to forecast fields. Unlike reanalysis fields, which are constructed after the fact, forecast fields are less accurate but available in real time. Can fate boundaries be computed from such imperfect fields, and can they better reveal a storm's possible tracks than operational ensembles?

Consider Khanun as an example. Khanun formed on 26 July 2023 (*31*). In forecasts issued on or before 29 July, all members of the European Centre for Medium-Range Weather Forecasts (ECMWF) ensemble (*32*, *33*) predicted a track toward China with small spread (Fig. 5A, inset). China prepared for landfall, and mainland Japan did not (*34*, *35*). On 30 July, a second branch appeared in the ensemble, pointing toward Japan (Fig. 5B, inset). Agencies in Japan rapidly began preparations, and an official warning was issued for Japan on 1 August (*36*). Khanun recurved and followed the second branch. The second branch was thus the critical signal, and the earlier it is predicted, the more time there is to prepare.

To see whether it could have been predicted earlier, we use only what was available on each day, namely the ECMWF deterministic forecast issued that day. We compute directional FTLE for the steering flow of the 29 and 30 July forecasts. In both, a fate boundary separating straight-moving from recurving trajectories lies close to Khanun (Fig. 5, A and B). Khanun sits on the China side, consistent with the deterministic tracks pointing toward China (thick purple lines in the insets). The recurving branch is nonetheless present in both fields, including that of 29 July, when the ensemble contained no recurving member. The recurving branch is further found as early as the 27 July forecast (fig. S10). The fate boundary thus predicts the bifurcation three days before the ensemble, using only what forecasters already had.

**Discussion**

In summary, fate boundaries provide a mechanistic and predictive framework for track bifurcation. Because it is in effect an exhaustive search over trajectories of a given steering flow, it can give warnings of alternative branches earlier than ensembles (Fig. 5), and even reveal fates entirely missed by ensembles (Figs. 3D and 4F and fig. S9). By applying the decades-old concept of Lagrangian coherent structures to tropical cyclone track forecasting, we have shed new light on the longstanding puzzle of track bifurcation.

For operational forecasting, we suggest computing fate boundaries routinely in real time. When a storm lies close to a boundary, forecasters should flag a qualitatively different track, even if the ensemble forecast shows no sign of it. For operational ensembles to better sample possible tracks, we propose that, rather than simply enlarging the ensemble, its members be deliberately seeded on both sides of a fate boundary.

Finally, we note that fate boundaries often separate devastating from benign tracks of storms that begin only slightly apart, as for Khanun in Fig. 3D and storms in Fig. 4, A and D. This hints at an application beyond prediction. Should it ever become technologically feasible to nudge a storm's position, the fate boundaries specify where, when, and in which direction to apply the nudge for maximum leverage, and how little would suffice. Whether such interventions should ever be exercised is a separate question (*37*, *38*).

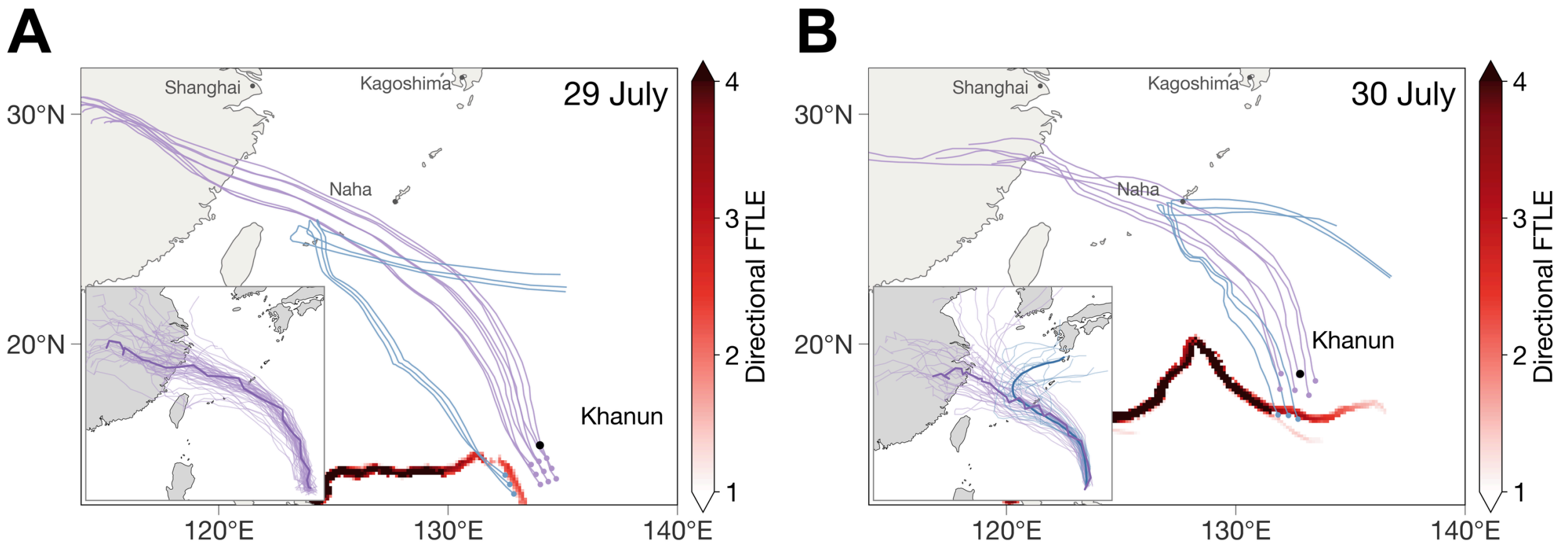

**Fig. 5. Khanun's fate boundaries in the forecast atmosphere.** As in Fig. 3, but with the steering flow taken from ECMWF deterministic forecasts initialized at 00:00 UTC 29 July (**A**) and 30 July 2023 (**B**), and with tracers advected from the forecast initial time to 00:00 UTC 6 August 2023 (SM, materials and methods). The black dot marks Khanun's observed position at the forecast initial time; insets show the 168-h ensemble forecast, colored by branch, with the deterministic forecast (thick purple) and, in (B), the mean track of the recurving members (thick blue). The fate boundary nearest to Khanun is shown, with tracer pairs seeded on either side of it.

**Funding:**
L. Li acknowledges support from the National Natural Science Foundation of China through the Excellent Young Scientists Fund Program (Overseas) and the General Program (Grant No. 12572474), and from the Fundamental Research Funds for the Central Universities of China. P.C. acknowledges support from the Okinawa Institute of Science and Technology Graduate University.

**Data, code, and materials availability:**
The ERA5 reanalysis data on pressure levels and single levels are available from the Climate Data Store of the Copernicus Climate Change Service (*39*, *40*). Tropical-cyclone best-track data are available from the International Best Track Archive for Climate Stewardship (IBTrACS) project (*41*). TIGGE ensemble and high-resolution forecast fields are available from the ECMWF Data Store (*42*), and the corresponding tropical-cyclone track products from the NSF NCAR Geoscience Data Exchange (*43*). Source data for the directional FTLE fields and tracer trajectories of all five cases, and additionally the FTLE fields for Khanun, together with Python scripts that reproduce the panels from these data, will be made publicly available upon publication.

The code for the minimal model including its tracer trajectory integration and the forward FTLE computation will be made publicly available upon publication.

Supplementary Materials for

# Solving the puzzle of tropical cyclone track bifurcation

Wei Zou, Jizhou Li, Xurui He, Lei Liu, Pinaki Chakraborty & Lin Li

Corresponding author: Lin Li, lin.li@scu.edu.cn

**The PDF file includes:**

## Materials and Methods

### Track and atmospheric data

Best-track positions of tropical cyclones are obtained from the International Best Track Archive for Climate Stewardship (IBTrACS v04r01) (*41*, *44*). Forecast positions are obtained from eight centers: the European Centre for Medium-Range Weather Forecasts (ECMWF), the National Centers for Environmental Prediction (NCEP), the Met Office, the Japan Meteorological Agency (JMA), Météo-France, the Korea Meteorological Administration (KMA), the China Meteorological Administration (CMA) and the Canadian Meteorological Centre (CMC). These are taken from the THORPEX Interactive Grand Global Ensemble (TIGGE) archive (*33*), supplemented by the Automated Tropical Cyclone Forecasting (ATCF) archive (*45*) for CMC and some NCEP tracks that are missing from TIGGE.

Reanalysis fields of atmosphere are taken from the fifth-generation ECMWF atmospheric reanalysis (ERA5) (*26*). We use mean sea-level pressure, the horizontal wind and geopotential on 16 pressure levels (1,000 and 925 hPa, and every 50 hPa from 850 to 200 hPa), sampled every 3 h from the hourly ERA5 single-level and pressure-level products on a 0.25° grid. Forecast fields of atmosphere are taken from the ECMWF high-resolution deterministic forecast (HRES) (*32*, *33*). We use mean sea-level pressure, the horizontal wind and geopotential height on eight pressure levels (1,000, 925, 850, 700, 500, 300, 250 and 200 hPa), retrieved at 6-h intervals on a 0.25° grid.

### Minimal point-vortex model

We construct the minimal model with two point vortices held at fixed positions, one to the northwest and the other to the southeast, 2,000 km apart (half-separation $l_0 = 1{,}000$ km), representative of the separation between the continental and oceanic highs around Khanun's turn (~1,900–2,600 km). Each vortex carries a clockwise (anticyclonic) circulation $\Gamma_i(t) < 0$ and induces an azimuthal flow of magnitude $|\Gamma_i|r/[2\pi(r^2 + a^2)]$ at distance $r$ from its center; the core radius $a = 200$ km keeps the velocity finite near the center (Scully regularization (*46*)). The velocity at any point is the vector sum of the two vortex flows, and storms are advected as passive point tracers. The trajectory is integrated with a fourth-order Runge–Kutta scheme (time step 30 s).

Both circulations begin equal in magnitude, $|\Gamma_i| = \Gamma_0 = 9.312 \times 10^7$ m$^2$ s$^{-1}$, corresponding to a tangential speed of ~14 m s$^{-1}$ at distance $l_0$. Over hours 8–24 of the 48-h integration, their magnitudes evolve linearly to final values of $0.6\Gamma_0$ (northwestern vortex) and $1.4\Gamma_0$ (southeastern vortex), and are held constant otherwise. We find that the time evolution of the circulations is necessary for the acute-angle turn of the tracer path.

We express lengths in units of the half-separation $l_0$ = 1,000 km, time in units of $\tau_0$ = 16 h, and velocity in units of $u_0 = l_0/\tau_0 \approx 17$ m s$^{-1}$. In these units, the circulation magnitude is $\Gamma_0\tau_0/l_0^2 \approx 5.4$ and the core radius $a/l_0 = 0.2$; the 48-h window spans $t/\tau_0$ = 0–3 and the ramp $t/\tau_0$ = 0.5–1.5.

### Ensemble-track bifurcation screening

To identify tropical cyclones with a strong signature of bifurcation in their predicted tracks, we screen the 1,179 Northern Hemisphere tropical cyclones recorded in IBTrACS from 2011 to 2025. For each initial time, we pool the 120-h forecasts from all centers into a single ensemble and retain only those with at least 20 members, so that the ensemble is large enough to resolve two distinct branches. To exclude weak storms, the observed 1-min maximum sustained wind must reach hurricane strength (32.9 m $s^{-1}$) within these 120 h and must not decay below tropical-storm intensity (17.5 m $s^{-1}$) at any time during this period. We then apply two criteria to identify a bifurcation: (1) the members form two sufficiently separated clusters, and (2) the two clusters diverge in direction, rather than merely separating in distance while moving in the same direction.

For the first criterion, the members' 120-h positions are divided into two clusters by k-medoids clustering (*47*) on great-circle distance, initialized with the two members farthest apart. For the split to be considered sufficiently separated, the mean silhouette width (*48*) must be at least 0.5. We further require that the smaller cluster contain at least 20% of the members, and that at least two forecasting centers each contribute at least two members to both clusters, so that the split does not rest on a single center.

For the second criterion, we first represent each branch by its mean track, and then calculate its heading. The heading of a mean track at a given lead time is defined as the direction from its position 24 h earlier to its current position. The heading difference between the two branches is computed every 6 h from 24 to 120 h, excluding lead times at which either branch moves less than 50 km in 24 h, and we require that it reach 60° at least once. The branches are further required to separate across the track rather than merely along it. The along-track direction is defined as the direction from the 72-h to the 120-h position of the all-member mean track, and the line joining the 120-h positions of the two branch-mean tracks must extend at least 600 km perpendicular to this direction.

A tropical cyclone is selected if both criteria are met for at least one initial time. Of the 1,179 tropical cyclones screened, 61 are selected. We additionally include Harvey (2017) for its devastating impact (Fig. 4B) and four cases from the 2026 season (fig. S8).

**Tropical-cyclone center identification**

The seeded tropical-cyclone trajectories are computed in the steering flow of the reanalysis and forecast fields, which requires removing the storm's own circulation from those fields. The removal begins with locating the storm center. A storm's centers at different heights, however, rarely coincide (the vortex can be tilted), so we identify the centers from the surface up through the pressure levels.

We locate the surface center with an objective algorithm adapted from the GFDL vortex tracker of Marchok (*49*), run operationally at NCEP. The search is initialized at a first-guess position: the IBTrACS best-track position nearest in time. Centered on this first-guess position, we search the mean sea-level pressure and geopotential heights at 850 and 700 hPa for their storm-scale minimum, and define the surface center as the mean of the three (the average is steadier than any single center). Each minimum is located by an iterative Barnes analysis (*50*): within a square box centered on the guess position, initially 600 km in half-width, the field is smoothed with a Gaussian kernel with an e-folding radius of 75 km to suppress sub-storm-scale noise, and the minimum of the smoothed field is taken as an updated center; the search is repeated five times, each subsequent pass re-centers on the previous estimate and halves its width, from 600 km down to 37.5 km. In parallel, the mesh spacing is halved at each iteration, from the native 0.25° grid down to about 1.7 km, so the minimum is located at sub-grid precision. A minimum is accepted only if it lies within 275 km of the first-guess position.

Centers in the forecast fields are located in the same way, except for the first guess: at long lead times, the forecast storm can drift far from the best track, so if any of the three minima is rejected, the search is seeded from the center at the preceding forecast time.

We then locate a center on each pressure level independently, to follow the vortex as it tilts with height. On each level, the center is taken as the maximum of the cyclonic relative vorticity, smoothed with a Gaussian kernel with a standard deviation of 50 km to suppress grid-scale noise. The maximum is sought within a ±4° box centered on the surface center, and accepted only if it lies within 200 km of that center; the wide search box and the acceptance radius together ensure that the maximum belongs to the storm rather than to a neighboring system. A level whose maximum is rejected is assigned a center by linear interpolation in pressure between the nearest accepted levels above and below; a rejected level with no accepted level above or below it takes the center of the nearest accepted level.

**Tropical-cyclone vortex removal**

We remove the storm circulation independently at each pressure level (*51*, *52*), so that the removal follows the vortex as it tilts and varies in size with height. The procedure combines two established tools: the kinematic vortex filter of the GFDL system to delimit the storm (*53*, *54*), and the vorticity–divergence inversion of Galarneau (*55*) and Davis (*56*) to subtract its winds.

We first delimit the storm with a kinematic filter. The filter acts on the disturbance wind, the total wind minus its area-weighted mean over a box of 5.5° half-width (approximately 600 km) centered on the level's center. This disturbance field is expressed in storm-centered polar coordinates, with the azimuth divided into 24 sectors of 15°. In each sector we search radially outward from the radius of maximum tangential wind for the filter radius $r_f(\theta)$: the first radius at which the disturbance tangential wind falls below 3 m s$^{-1}$, or the second radius at which it falls below 6 m s$^{-1}$ where its radial decay had already weakened ($-\partial v_\theta/\partial r < 4 \times 10^{-6}$ s$^{-1}$). The vortex radius is then $r_0(\theta) = \min[1.25 r_f(\theta), R]$, reduced to any smaller radius at which the tangential wind reversed sign, marking the outer edge of the storm. The cap $R$ is an upper bound on the removal radius, which prevents the filter from expanding into the surrounding flow; it is estimated jointly with the steering layer and the β drift along the track (see "Steering-flow estimation and trajectory computation"). The 24 radii $r_0(\theta)$ define a 24-sided polygon that encloses the region attributed to the storm.

Within this polygon, we isolate and subtract the storm's circulation following the inversion approach. The relative vorticity $\zeta = \partial v/\partial x - \partial u/\partial y$ and horizontal divergence $\delta = \partial u/\partial x + \partial v/\partial y$ are computed from the total wind. Inside the polygon, where $\zeta$ and $\delta$ are dominated by the storm, we attribute both entirely to the storm (*55*, *56*). The vortex sources $\zeta_{tc}$ and $\delta_{tc}$ are therefore $\zeta$ and $\delta$ inside the polygon and zero outside, tapered to zero over the outermost 10% of the vortex radius by a half-cosine weight, so that neither source is cut off abruptly at the edge. Inverting the two sources gives the vortex streamfunction and velocity potential: $\nabla^2\psi_{tc} = \zeta_{tc}$ and $\nabla^2\varphi_{tc} = \delta_{tc}$. The Poisson equations are solved by $N$ Jacobi iterations from a zero initial field, which confine the vortex-induced wind to scales below $L$ = 1,200 km (from $L = h\sqrt{(N/2)}$, where $h$ is the grid spacing of the inversion and $N$ the number of iterations; $N \approx$ 4,260 for $h \approx$ 26 km), a scale comparable to the maximum filter radius in the vortex-removal scheme of Kurihara et al. (*54*), beyond which environmental rather than storm wind would be removed.

**Steering-flow estimation and trajectory computation**

For the atmospheric fields, trajectories are computed not from the steering alone but with a beta-advection model (*28*), where the storm moves with its environmental steering plus a β drift, typically toward northwest in the Northern Hemisphere. The tracer velocity is $\mathbf{V} = \mathbf{V}_{\text{steering}} + \mathbf{V}_{\beta}$. The steering $\mathbf{V}_{\text{steering}}$ is the pressure-weighted deep-layer mean of the vortex-removed horizontal wind between a fixed base at 850 hPa and a top $p$, where the vortex is identified in, and removed from, the atmospheric fields (see "Tropical-cyclone vortex removal") with the removal radius capped at $R$. The β drift $\mathbf{V}_{\beta} = (-\beta_{\text{W}}, \beta_{\text{P}})$ is a constant vector, added directly to the tracer velocity. The depth of the layer that steers a cyclone varies with the storm's own vertical extent (*55*, *57*, *58*) and hence changes as the storm develops, and the appropriate removal radius and β drift similarly vary at different time. We therefore estimate these four quantities jointly along the track.

For each starting time, taken every 3 h along the observed track, we seek the parameter vector $\boldsymbol{\theta}_i = (R_i, p_i, \beta_{\text{W},i}, \beta_{\text{P},i})$ that best reproduces the observed motion over a finite window, typically 120 h. Within this window the parameters are held fixed while the steering field evolves with the 3-hourly ERA5 data. $R$ is searched from 400 to 800 km in 50-km steps, $p$ from 500 to 200 hPa in 50-hPa steps, and $\beta_{\text{W}}$ and $\beta_{\text{P}}$ each from 0 to 2 m s$^{-1}$ in 0.1 m s$^{-1}$ steps (27,783 combinations). For each combination a tracer is released from the IBTrACS position and integrated on the sphere in longitude–latitude coordinates with a fourth-order Runge–Kutta scheme (time step 30 min), the wind being interpolated bilinearly in space and linearly in time between the 3-hourly fields. The misfit is $J_i(\boldsymbol{\theta}) = \sum_k [d_{i,k}(\boldsymbol{\theta}) / d_{\text{ref}}]^2$, where the sum runs over the $K_i$ IBTrACS fixes in the window and $d_{i,k}$ is the geodesic distance between the tracer and the IBTrACS position at the $k$-th verification time. A full 120-h window contains 40 verification times; where the track ends within 120 h, the integration stops at the last record and fewer terms are summed. The normalization by $d_{\text{ref}} = 324.9$ km, the mean 120-h track error of ECMWF forecasts (*59*), makes $J_i$ dimensionless. Because the misfit surface contains sharp, isolated minima that fit a particular track closely but are not robust to small parameter changes, selecting the single best combination is prone to overfitting. We therefore score each combination by the mean of $J_i$ over itself and its neighbors within one grid step in $R$ and $p$ and within 0.1 m s$^{-1}$ in each β component (up to 81 combinations), which favors flat minima over sharp ones (*60*) and substantially reduces overfitting. $J_i$ hereafter denotes this mean. Then the combination with the smallest mean is retained, and the search is repeated at the next starting position.

The above first pass treats each start time independently and gives the best discrete fit at each time, but the resulting parameters can nonphysically jump between neighboring times. To suppress such jumps, we then impose temporal continuity, by penalizing changes in the parameters between consecutive times, minimizing at each time,

$$F_i(\boldsymbol{\theta}_i) = J_i(\boldsymbol{\theta}_i) + \lambda\, D(\boldsymbol{\theta}_{i-1}, \boldsymbol{\theta}_i),$$

where

$$D(\boldsymbol{\theta}_{i-1}, \boldsymbol{\theta}_i) = (R_i - R_{i-1})^2/\sigma_R^2 + [H_i(p_i) - H_{i-1}(p_{i-1})]^2/\sigma_H^2 + (\beta_{\text{W},i} - \beta_{\text{W},i-1})^2/\sigma_\beta^2 + (\beta_{\text{P},i} - \beta_{\text{P},i-1})^2/\sigma_\beta^2$$

with $H_i(p)$ the geopotential height of pressure level $p$ at time $i$ at the storm position, and $\sigma_R = 50$ km, $\sigma_H = 1{,}000$ m and $\sigma_\beta = 0.3$ m s$^{-1}$ the scales against which changes in the parameters are measured. The value $\lambda = 0.2$ is a balance between the fit at individual times against the temporal continuity.

We then carry out the minimization iteratively. Starting from the first-pass optima, we evaluate $F_i$ for all combinations of $R$, $p$, and $\beta$ within ±100 km, ±100 hPa and ±0.6 m $s^{-1}$ of the current values, and take the combination with the smallest $F_i$ as the updated parameters. Restricting the candidates to this neighborhood prevents jumps to a distant combination with a similar misfit. The procedure is repeated until no parameter changes, with a maximum of 100 iterations.

The same workflow applies to operational forecast fields, with the tropical-cyclone track identified in the forecast fields replacing IBTrACS, and the storm's center and size determined from the forecast flow.

**Finite-time Lyapunov exponent (FTLE)**
The FTLE (*23*, *25*) is computed in the standard way from the flow map $\Phi$, which sends each initial position $\mathbf{x}_0$ at time $t_0$ to its position after a window $T$. Its spatial gradient $\nabla\Phi$ describes how an infinitesimal neighborhood of $\mathbf{x}_0$ is deformed by the flow. The eigenvalues of the right Cauchy–Green strain tensor $\mathbf{C} = (\nabla\Phi)^T\nabla\Phi$ are the squares of the factors by which that neighborhood is stretched along the principal directions of deformation, and the FTLE is the exponential rate implied by the largest eigenvalue $\lambda_{max}$: FTLE = $[1/(2T)] \ln \lambda_{max}$.

**Directional FTLE**
The standard FTLE yields high values wherever neighboring trajectories separate, whether through shear or through genuine divergence onto different paths (*24*), yet only the latter marks track bifurcations. We therefore construct a "directional FTLE" that responds only where neighboring tracers separate onto different headings, and remains insensitive where they part while preserving a common direction.

The construction has four steps. First, at each grid point $\mathbf{x}_0$ we advect a single tracer over the window $[t_0, t_0 + T]$ with the same flow map $\Phi$ used for the FTLE, and record its heading $\theta(\mathbf{x}_0)$: the direction of its net displacement from the start point $\mathbf{x}_0$ to the end point $\Phi(\mathbf{x}_0)$, measured counterclockwise from east.

Second, at each start point we compute the direction along which the heading changes most rapidly, given by the normalized gradient of the heading field, $\mathbf{n} = \nabla\theta/|\nabla\theta|$. (Where $\nabla\theta$ vanishes, or the tracer barely moves, the index defined below is left undefined.)

Third, we probe that direction with a pair. We seed two tracers a short distance $d_0$ apart ($d_0$ = 50 km), symmetrically about $\mathbf{x}_0$ along $\mathbf{n}$, with $\mathbf{x}^{\pm} = \mathbf{x}_0 \pm (d_0/2)\mathbf{n}$, so that the pair straddles the sharpest change of heading, and advect both over $[t_0, t_0 + T]$.

Fourth, from the advected pair we extract two dimensionless factors. The angle factor $|\Delta\theta| = |\theta^+ - \theta^-|$, where $\theta^+$ and $\theta^-$ are the headings at the endpoints of the two trajectories and the difference is wrapped to $[0, \pi]$ (in radians), measures how differently the flow steers the pair. The separation factor, $[\ln(d_{min}/d_0)]_+$, is built from $d_{min} = \min(d^+, d^-)$, where $d^+$ is the closest geodesic distance from the endpoint of the trajectory of $\mathbf{x}^+$ to the entire trajectory of $\mathbf{x}^-$, and $d^-$ is defined symmetrically (fig. S3); it measures whether the paths genuinely part rather than merely slide along one another.

The positive part $[\cdot]_+ = \max(\cdot, 0)$ holds the factor at zero until the pair separates beyond its initial spacing $d_0$.

The directional FTLE is the product of the two factors, $|\Delta\theta|\,[\ln(d_{\min}/d_0)]_+$. Under shear, both factors stay small: the trajectories remain close (small $d_{\min}$) and the headings nearly parallel (small $|\Delta\theta|$). For genuinely divergent trajectories, by contrast, both grow large (fig. S3). In principle, either factor alone can signal bifurcation; in practice their product performs better. We do not, however, divide by the window length $T$ to form a rate, as is done for the FTLE. There, the logarithm of the stretching grows in proportion to $T$, so the quotient converges to a well-defined rate, whereas the angle between two headings is bounded by $\pi$ and does not accumulate with time. The directional FTLE is therefore dimensionless.

Note that the heading is an Earth-frame quantity, so the directional FTLE, unlike the FTLE, is not frame-invariant. This is deliberate. Where a tropical cyclone goes on the Earth is itself defined in the Earth frame.

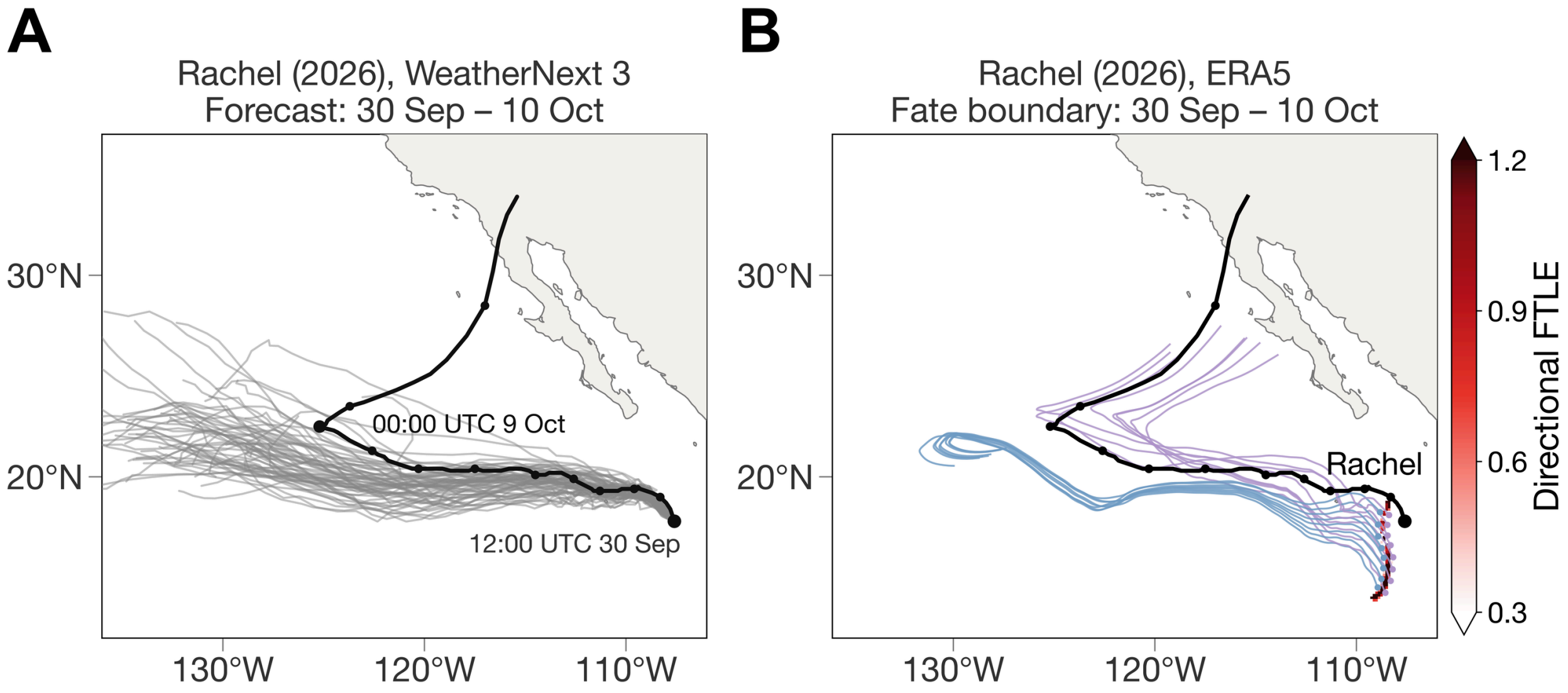

**Fig. S1. Bifurcation of Rachel (2026) not captured by Google DeepMind's WeatherNext 3.**

(**A**) 240-h WeatherNext 3 ensemble forecast from 12:00 UTC 30 September (thin gray lines, 64 members). Thick black line, observed track; dots every 24 h. (**B**) Directional FTLE (red shading) from ERA5, computed over 240 h from 12:00 UTC 30 September. Only the fate boundary relevant to the storm is shown; tracers seeded on either side of it (purple, storm's side; blue, opposite side) are advected over the same window. Black line, observed track.

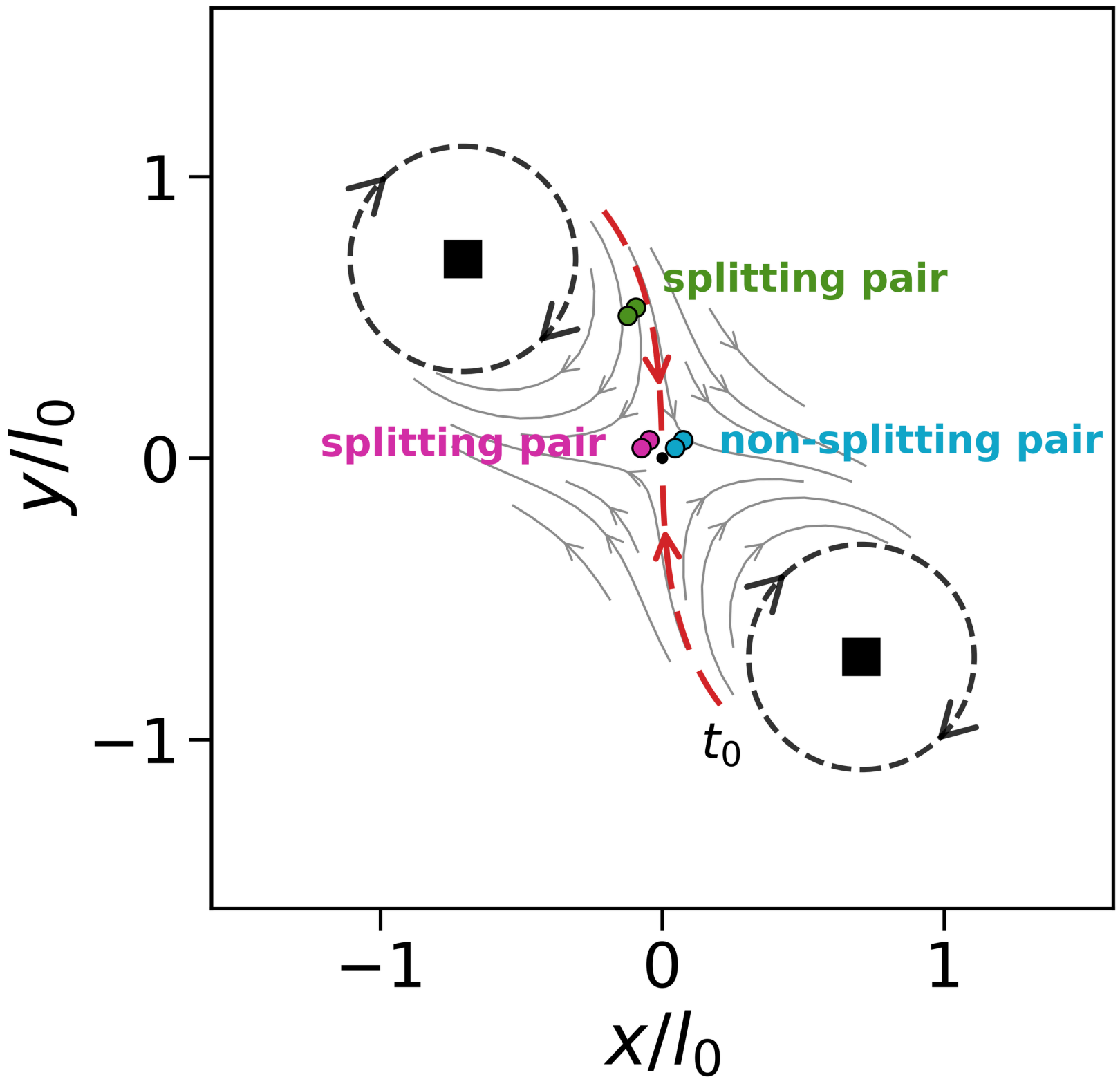

**Fig. S2. Fate boundary and axis of contraction.** The closest antecedent of the fate boundary in the tropical-cyclone literature is the "axis of contraction" (*19*) of a deformation steering flow. At a given instant, the axis of contraction is the streamline along which the flow converges toward the stagnation point. It is thus an instantaneous feature, whereas the fate boundary is a finite-time feature, defined over a time window. In a steady flow the two coincide, but the atmosphere is rarely steady. Even in the minimal model, the axis of contraction at t = 0 (red dashed line) fails to separate the members of either splitting pair.

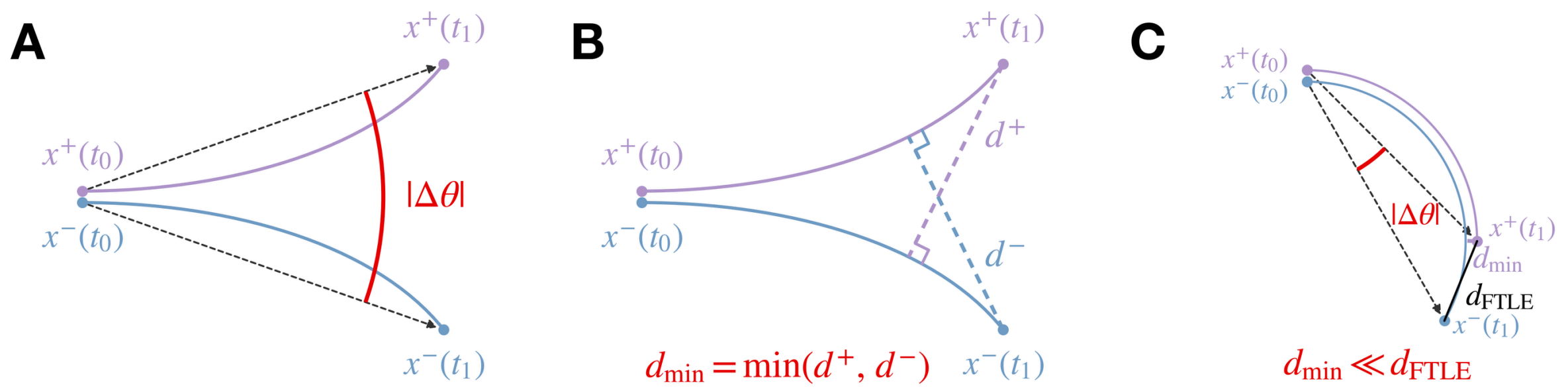

**Fig. S3. Definition of the directional FTLE.** The directional FTLE identifies where the steering flow routes neighboring tracers onto divergent paths (see materials and methods), and is defined as $|\Delta\theta|\ [\ln(d_{\min}/d_0)]_+$, the product of (A) the angle between their net displacements and (B) the separation reached by their trajectories, where the positive part $[\cdot]_+$ keeps this factor at zero until the pair separates beyond its initial spacing $d_0$. In each panel, two tracers begin a short distance $d_0$ apart at $t_0$, $\mathbf{x}^+(t_0)$ (purple) and $\mathbf{x}^-(t_0)$ (blue), and are advected by the steering flow to $t_1$, reaching $\mathbf{x}^+(t_1)$ and $\mathbf{x}^-(t_1)$; black dashed arrows (A and C) show their net displacements. (**A**) $|\Delta\theta|$ (red arc) is the angle between the two headings (0–180°). (**B**) $d^+$ is the closest distance from the endpoint of the trajectory of $\mathbf{x}^+$ to the entire trajectory of $\mathbf{x}^-$, and $d^-$ is defined symmetrically; their smaller value, $d_{\min} = \min(d^+, d^-)$, sets the separation factor $[\ln(d_{\min}/d_0)]_+$. (**C**) The directional FTLE distinguishes shear from genuine divergence. Here, the shear carries the two endpoints apart, yielding a large separation, $d_{\text{FTLE}}$ (black line), yet the gap $d_{\min}$ stays narrow and the headings nearly parallel, holding the index low.

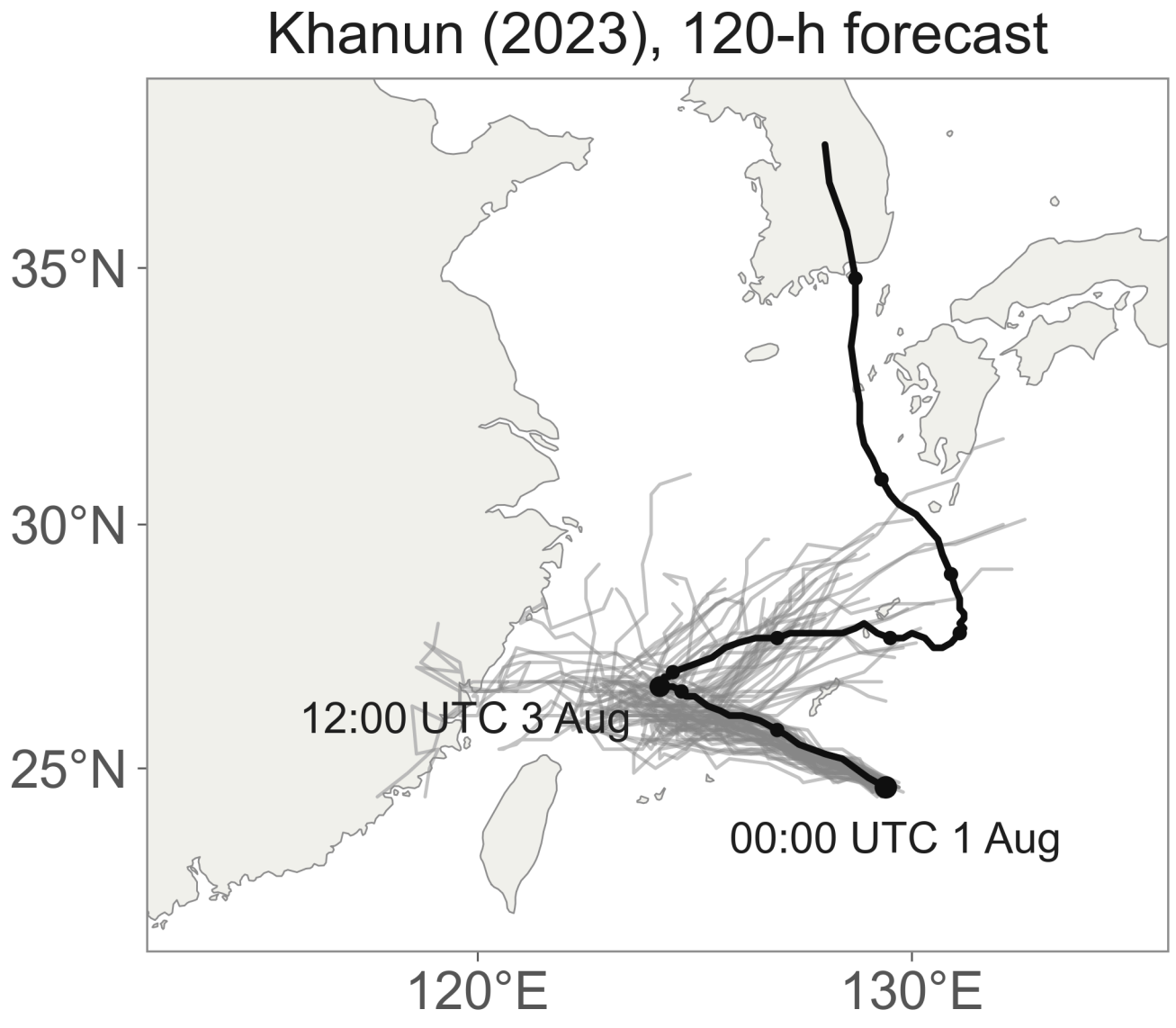

**Fig. S4. Ensemble forecast of Khanun (2023) issued at 00:00 UTC 1 August 2023.** As in Fig. 1D, thin gray lines are the 51 members of the ECMWF ensemble, and the thick black line is the observed track, with dots every 24 h.

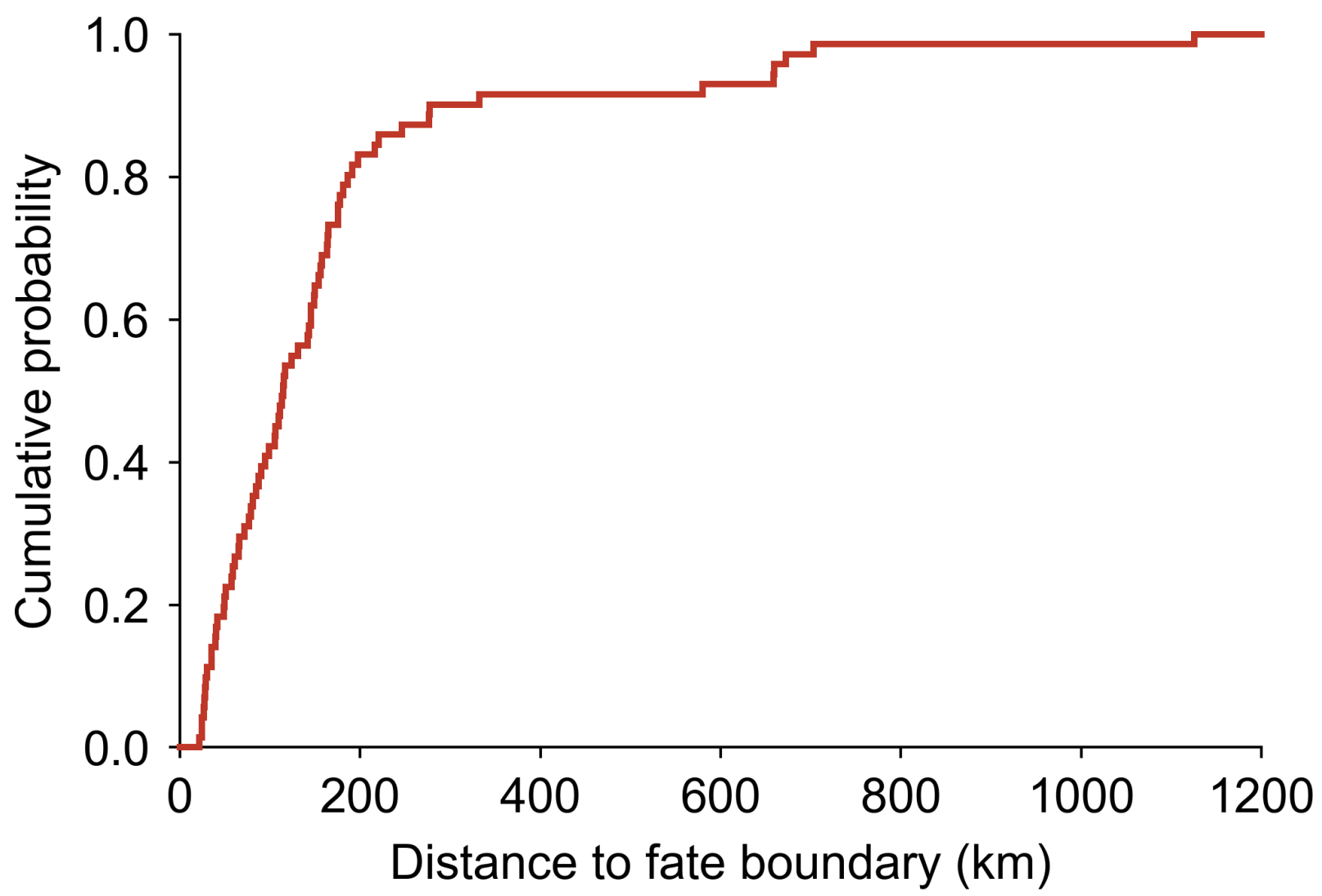

**Fig. S5. Cumulative distribution of storm-to-fate-boundary distance for all cases shown.** Distances are measured at the same time as shown in Fig. 4 and figs. S6 to S8.

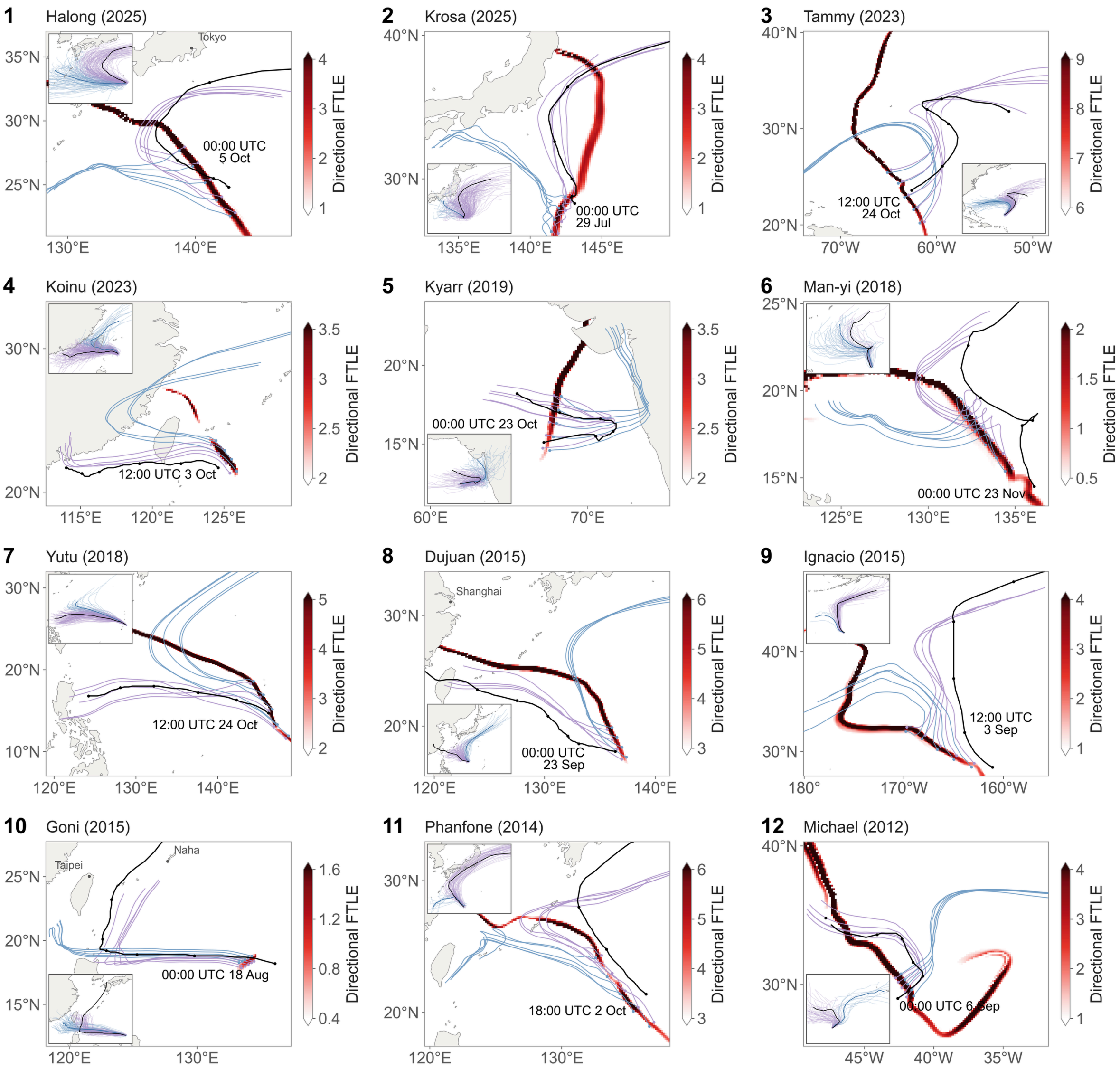

**Fig. S6. Fate boundaries in bifurcating tropical cyclone tracks.** Of 61 screened storms whose predicted tracks bifurcate, five are shown in Fig. 4 (alongside Harvey, which is not among the 61) and five that bifurcate twice are shown in fig. S7; the remaining 51 are shown here. Red shading is directional FTLE computed over 120 h from the start time labeled at the storm position (117, 114, and 84 h in panels 12, 20, and 28, where data end earlier). Only the fate boundary relevant to the bifurcation is shown; tracers seeded on either side of it (purple and blue) are advected over the same window. Black lines are observed tracks with dots every 24 h. Insets, multicenter 120-h ensemble forecasts, colored by branch; black lines, observed track; thick colored lines, mean track of each branch.

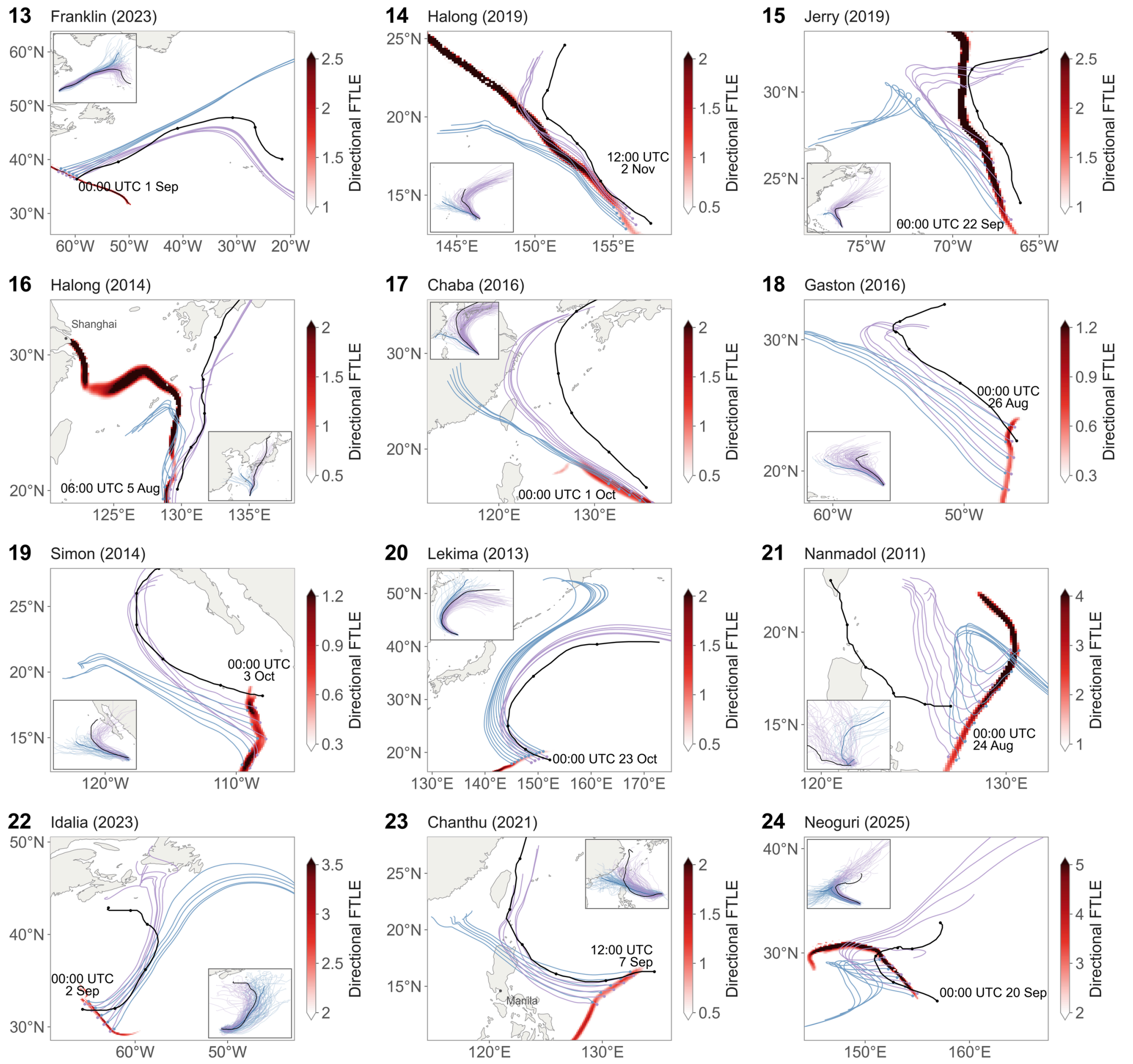

**Fig. S6.** (continued)

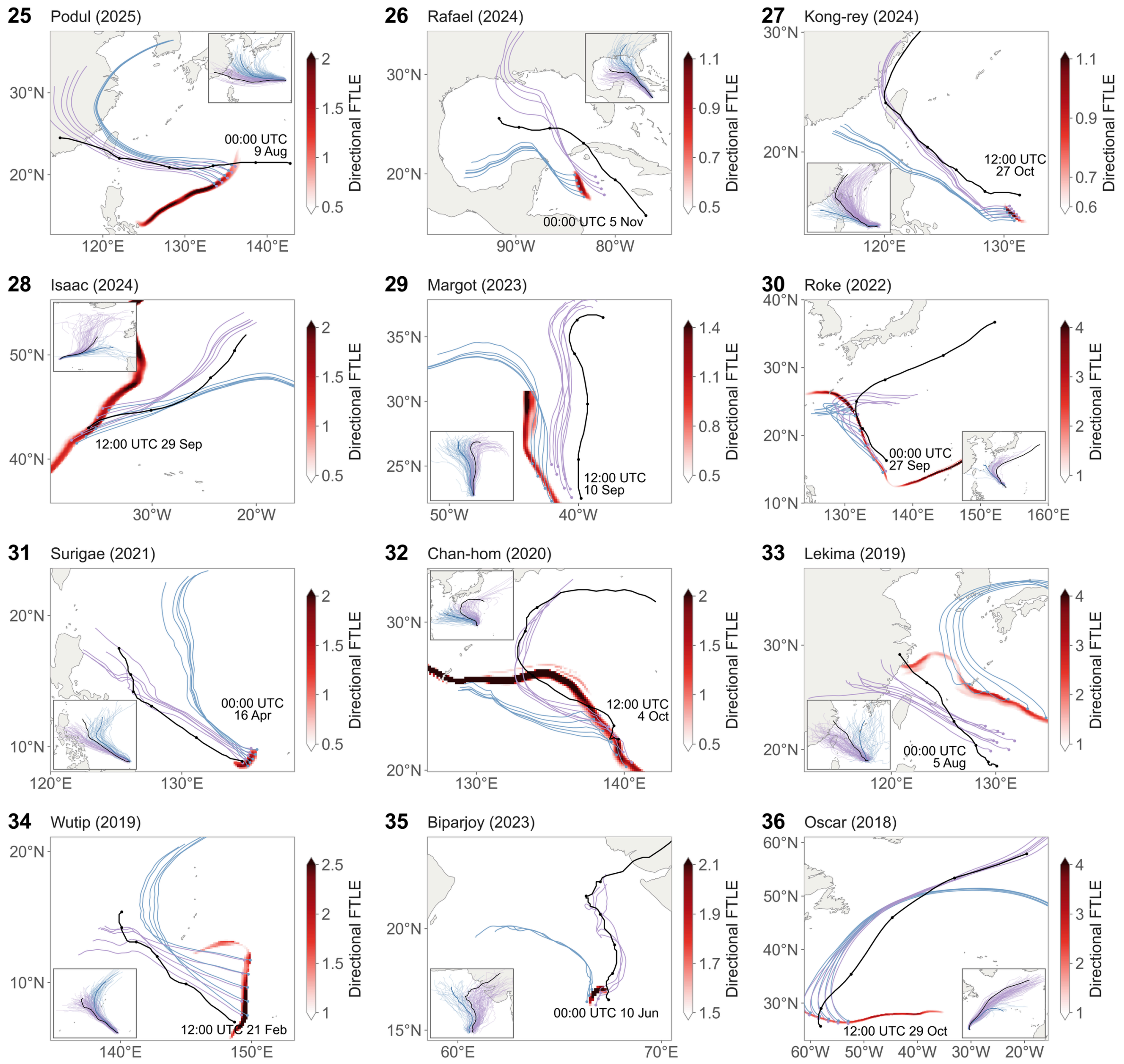

**Fig. S6.** (continued)

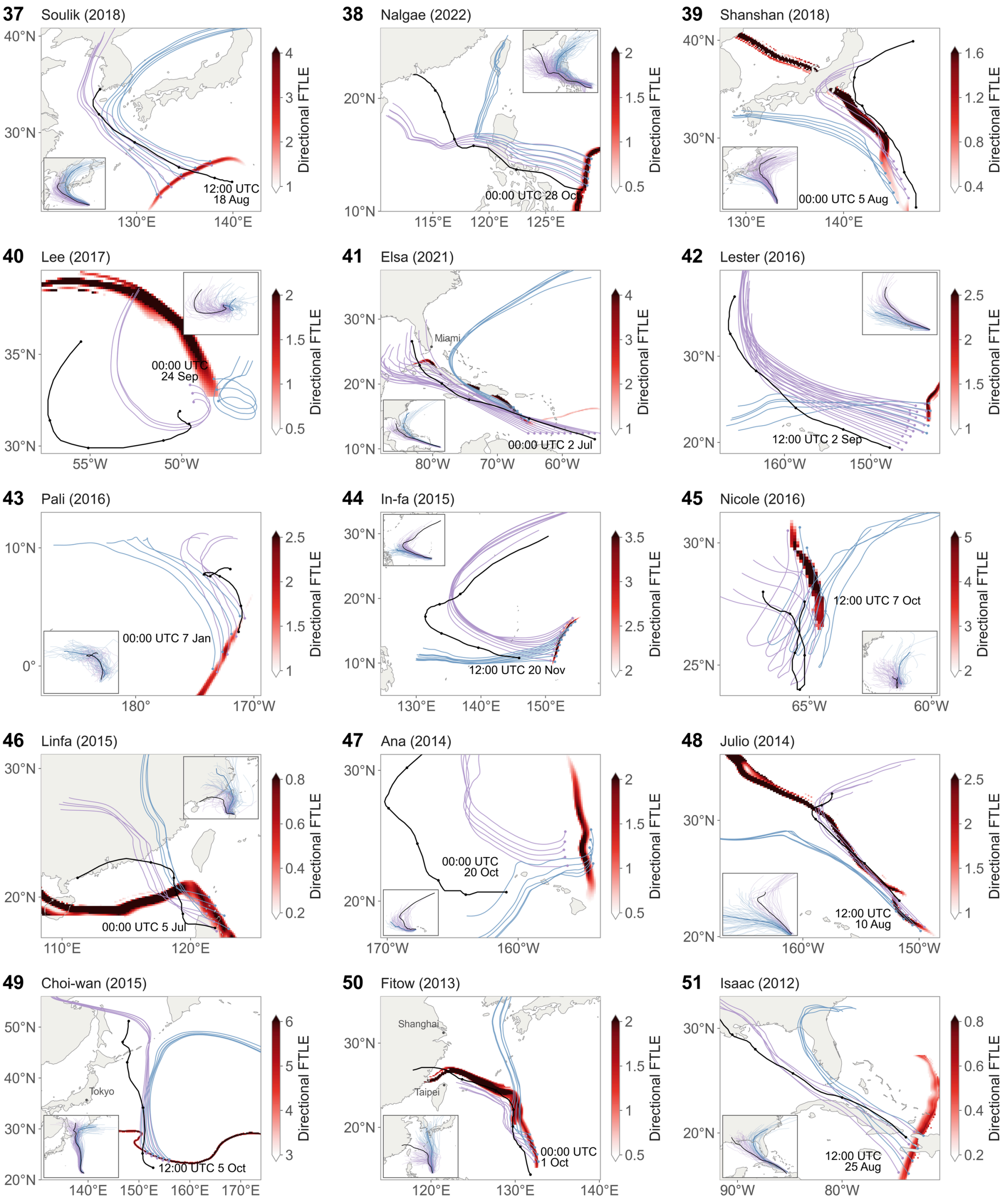

**Fig. S6.** (continued)

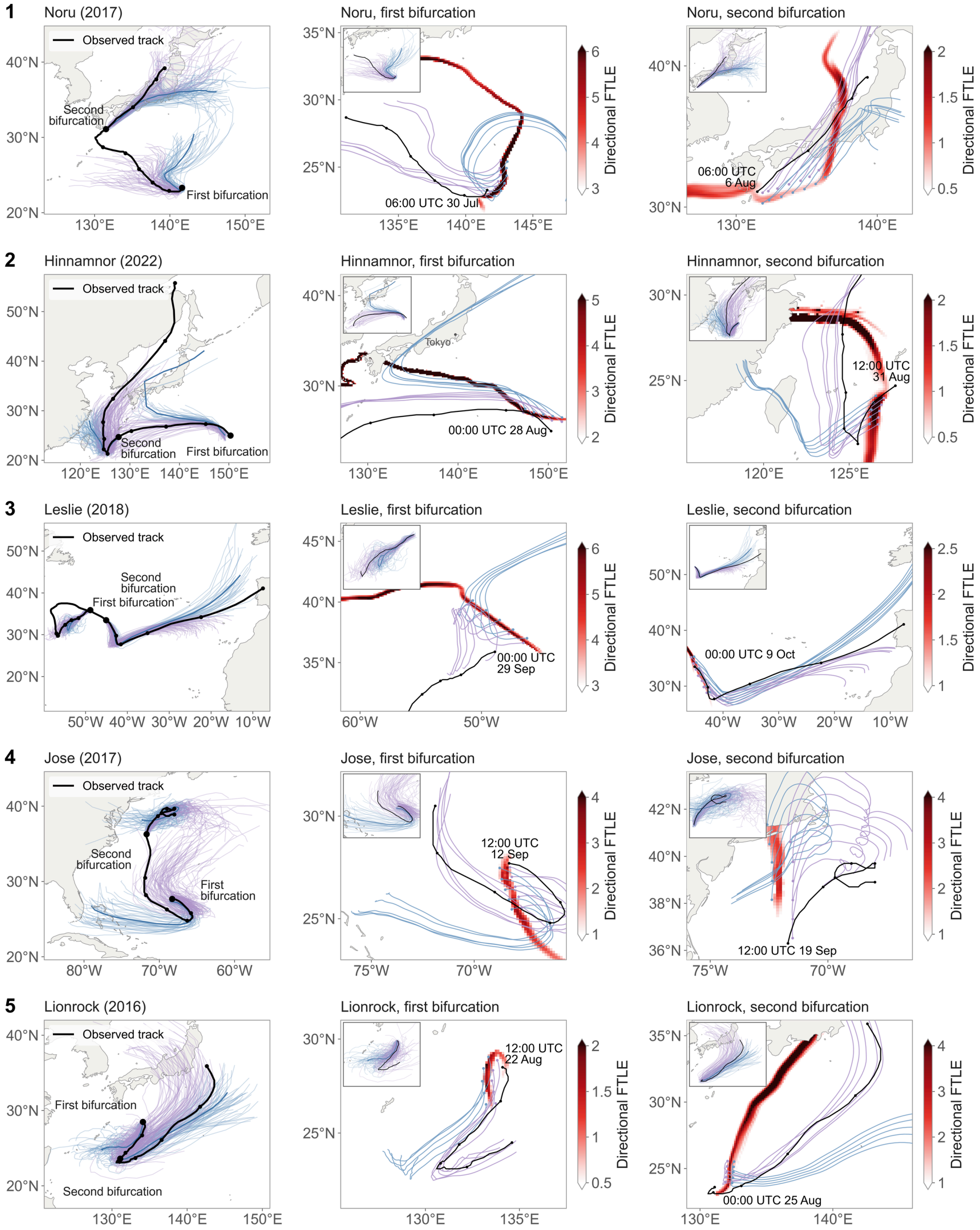

**Fig. S7. Fate boundaries in storms that bifurcate twice.** Left column: 120-h ensemble forecasts from the start times of the first and second bifurcations (large dots on the observed track), colored by branch (light purple, the branch followed by the observed track; light blue, the other); thicker lines are branch means. Middle and right columns: directional FTLE (red shading) at the first and second bifurcations, computed over 120 h from the start time labeled at the storm position (72 h for the second bifurcation of Noru, where the available data end earlier). Only the fate boundary nearest the storm is shown; tracers seeded on either side of it are advected over the same window: purple on the storm's side, blue on the other side. Black lines are observed tracks, with dots every 24 h. Insets as in fig. S6. (**1**) Noru (2017). (**2**) Hinnamnor (2022). (**3**) Leslie (2018). (**4**) Jose (2017). (**5**) Lionrock (2016).

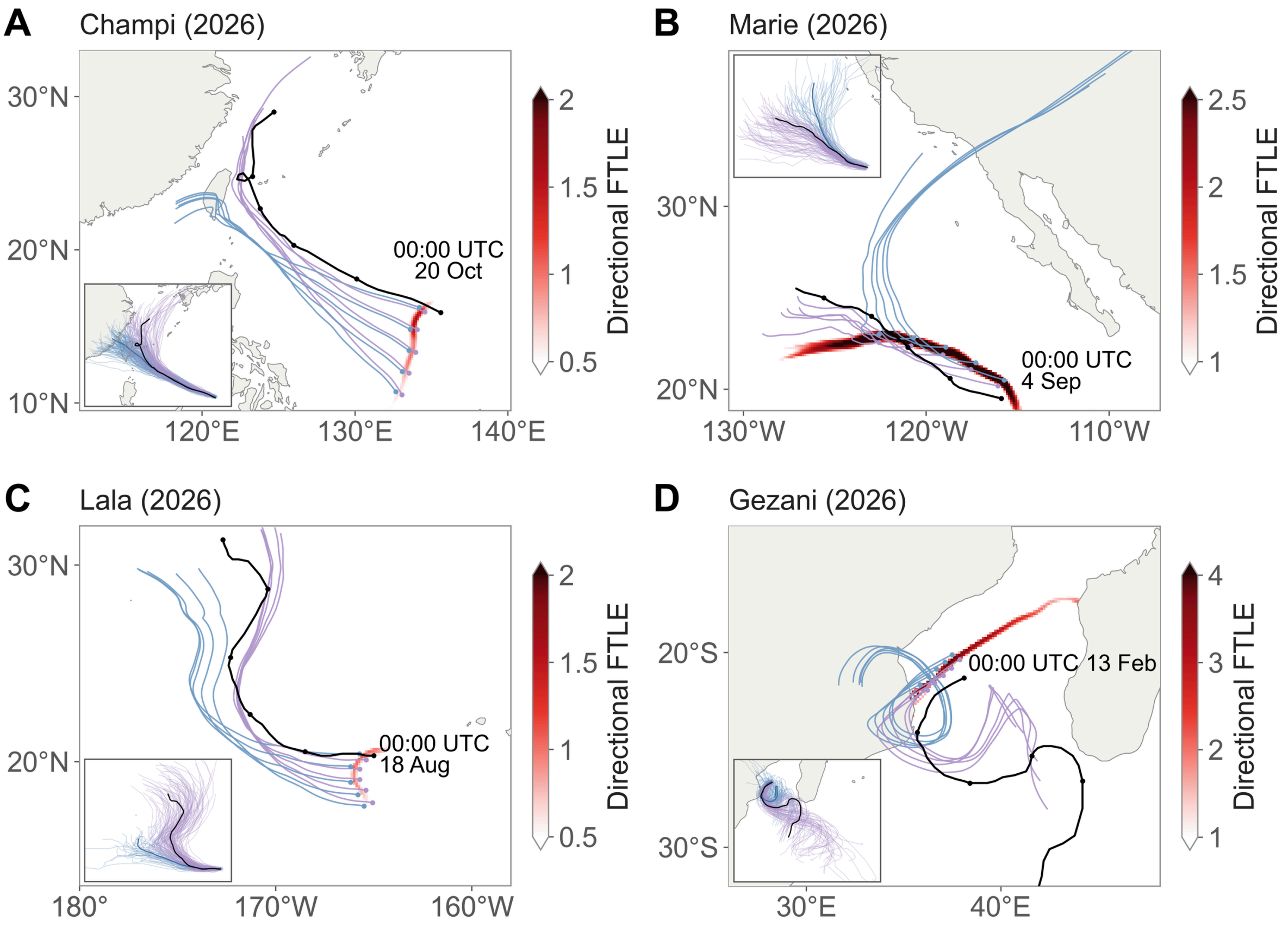

**Fig. S8. Bifurcating tropical cyclones of 2026.** We additionally analyzed cases from the 2026 season, which had not ended at the time of submission. (**A**) Champi (2026). (**B**) Marie (2026). (**C**) Lala (2026). (**D**) Gezani (2026), a Southern Hemisphere storm. Color conventions are as in Fig. 4.

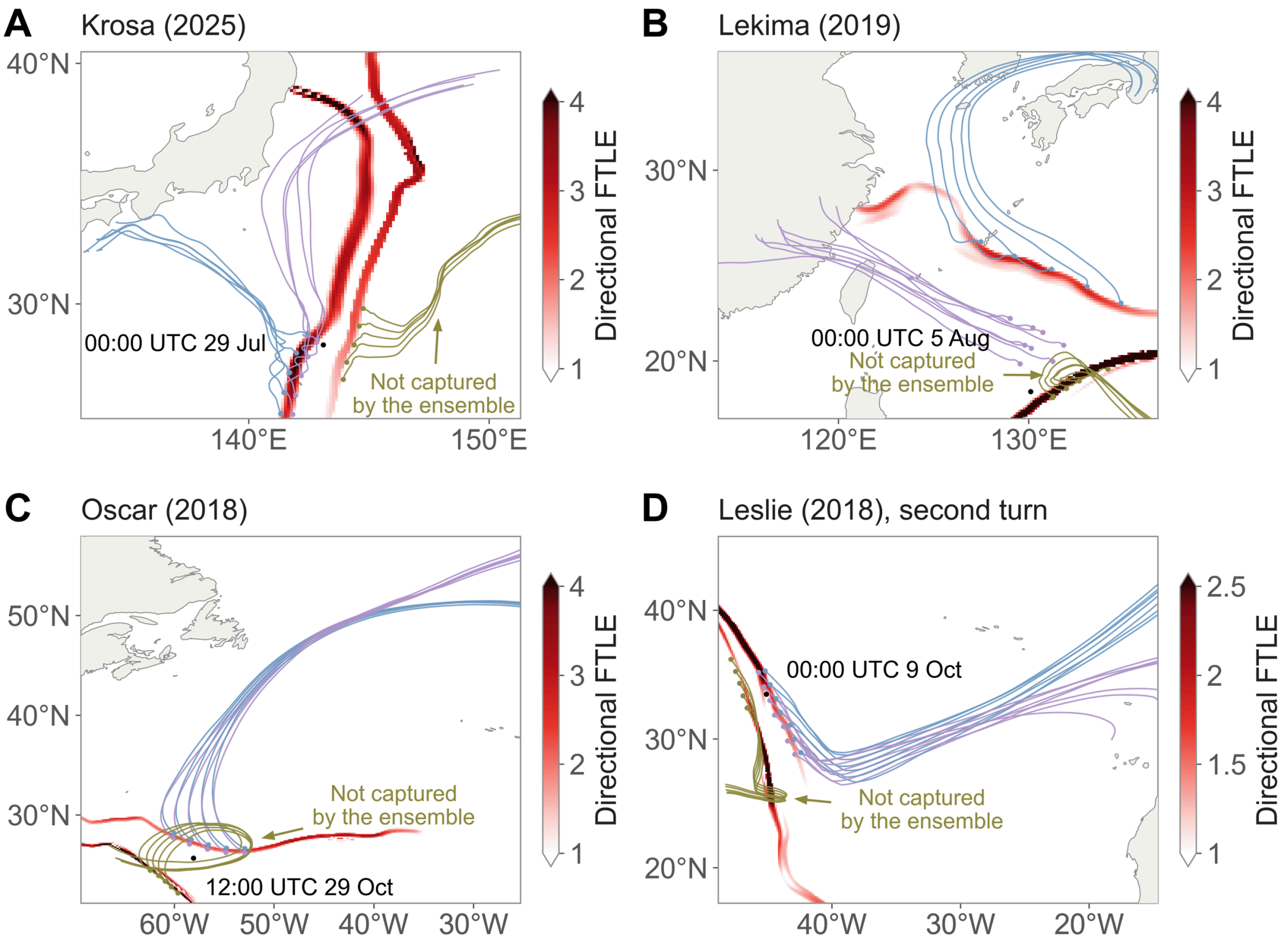

**Fig. S9. Fates missed by the forecast ensembles.** Directional FTLE (red shading) is computed over 120 h from the start time labeled at the storm position (black dot). Only the two ridges nearest the storm are shown. Purple and blue tracers are seeded on either side of the fate boundary shown in fig. S6 for (A) to (C) and in fig. S7 for (D): purple on the storm's side, blue on the other side. Olive tracers, seeded beside the other ridge, follow the branch not captured by the forecast ensembles (arrow). (**A**) Krosa (2025). (**B**) Lekima (2019). (**C**) Oscar (2018). (**D**) Leslie (2018), second turn.

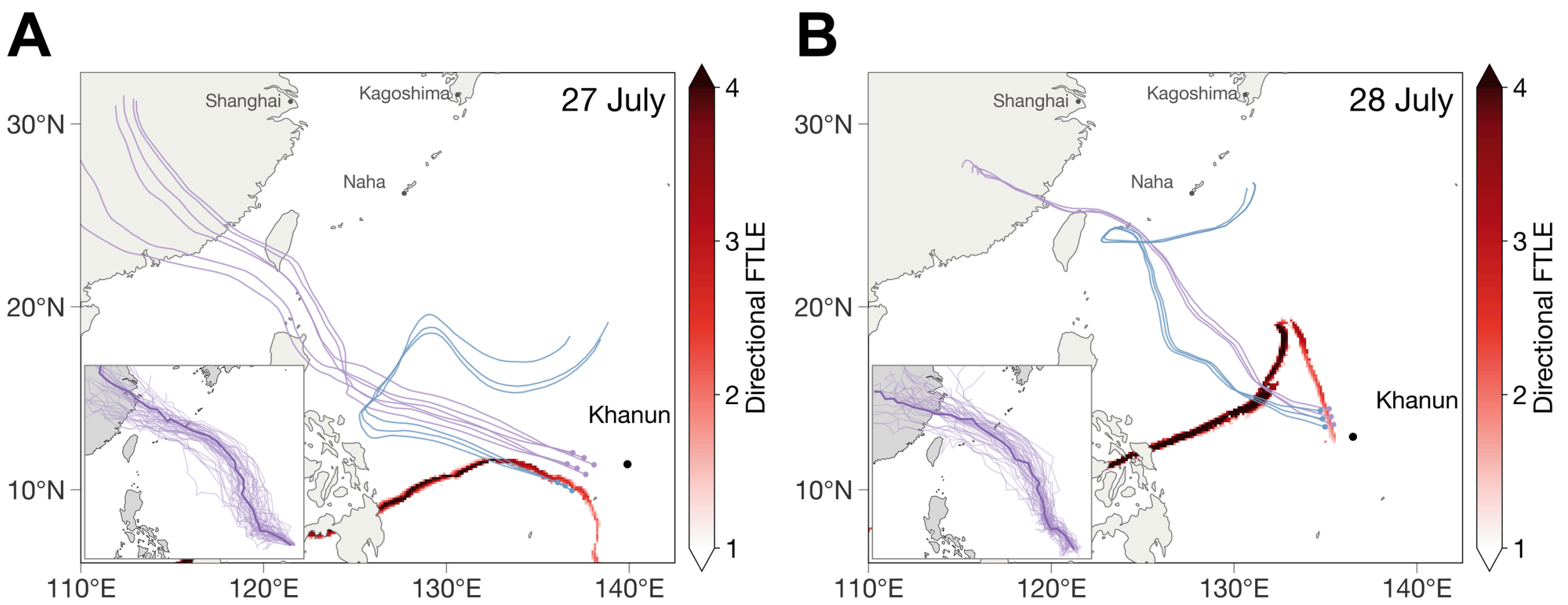

**Fig. S10. Fate boundaries of Khanun (2023) in the 27 and 28 July forecasts.** As in Fig. 5, but for ECMWF deterministic forecasts initialized at 00:00 UTC 27 July (**A**) and 28 July (**B**).